\documentclass{ws-procs975x65}

\begin{document}

\title{COLOR SUPERCONDUCTING QUARK MATTER IN \\ COMPACT STARS}

\author{D. B. BLASCHKE$^*$}

\address{Institute for Theoretical Physics, University 
of Wroclaw,\\% Max-Born pl. 9, 
50-204 Wroclaw, Poland\\
$^*$E-mail: blaschke@ift.uni.wroc.pl\\
www.ift.uni.wroc.pl/$\sim$blaschke}

\author{T. KL\"AHN}

\address{Theory Division, Argonne National Laboratory, \\
Argonne IL 60439-4843, USA\\
E-mail: Thomas.Klaehn@googlemail.com}

\author{F. SANDIN}

\address{Department of Physics, Lule{\aa} University of
Technology,\\
 97187 Lule{\aa}, Sweden\\
E-mail: Fredrik.Sandin@gmail.com}

\begin{abstract}
Recent indications for high neutron star masses ($M \sim 2 ~M_\odot$) and 
large radii ($R > 12$ km) could rule out soft equations of state and have 
provoked a debate whether the occurence of
quark matter in compact stars can be excluded as well.
We show that modern quantum field theoretical approaches to quark matter 
including color superconductivity and a vector meanfield allow a microscopic 
description of hybrid stars which fulfill the new, strong constraints.
For these objects color superconductivity turns out to be an essential 
ingredient for a
successful description of the cooling phenomenology in accordance with
recently developed tests.
We discuss the energy release in the neutrino untrapping transition as a new
aspect of the problem that hybrid stars masquerade themselves as neutron stars.
Quark matter searches in future generations of low-temperature/high-density 
nucleus-nucleus collision experiments such as low-energy RHIC and CBM @ FAIR
might face the same problem of an almost crossover behavior of the 
deconfinement transition. 
Therefore, diagnostic tools shall be derived from% (precursor) 
effects of color superconductivity. 
\end{abstract}


\bodymatter

\section{Introduction} %: modern compact star observations}
Recently, observations of compact stars have provided new data of high
accuracy which put strong constraints on the high-density behaviour of the
equation of state of strongly interacting matter otherwise not accessible in
terrestrial laboratories.
In particular, the high masses of $M=1.96 +0.09/-0.12~M_\odot$ and 
$M=2.73 \pm 0.25~M_\odot$ obtained in very recent measurements on the 
millisecond pulsars  PSR B1516+02B and PSR J1748-2021B, repectively 
\cite{Freire:2007jd}, together with the large radius of $R > 12$ km for the   
isolated neutron star RX J1856.5-3754 (shorthand: RX J1856)   
\cite{Trumper:2003we} point to a stiff equation of state at high densities.  
Measurements of high masses are also reported for compact stars in low-mass  
X-ray binaries (LMXBs) as, e.g.,  
$M=2.0\pm 0.1~M_\odot$ for the compact object in 4U 1636-536  
\cite{Barret:2005wd}. 
For another LMXB, EXO 0748-676,   
constraints for the mass $M\ge 2.10\pm 0.28~M_\odot$  
{\it and} the radius $R \ge 13.8 \pm 0.18$ km %for the same object   
have been derived \cite{Ozel:2006km}.  
The status of these data is, however, unclear since the observation of a   
gravitational redshift $z=0.35$ in the X-ray burst spectra 
\cite{Cottam:2002cu} 
could not be confirmed thereafter despite numerous attempts 
\cite{Cottam:2007cd}.

Measurements of rotation periods below $\sim 1$ ms as discussed for 
XTE J1739-285 \cite{Kaaret:2006gr}, on the other hand, would disfavor too 
large objects corresponding to a stiff EoS and would thus leave only a tiny 
window of very massive stars in the mass-radius plane 
\cite{Lavagetto:2006ew,Bejger:2006hn}
for a theory of compact star matter to fulfill all 
above mentioned constraints. 
\begin{figure} [ht]
\begin{tabular}{ll}
\includegraphics[angle=270,width=0.5\textwidth]{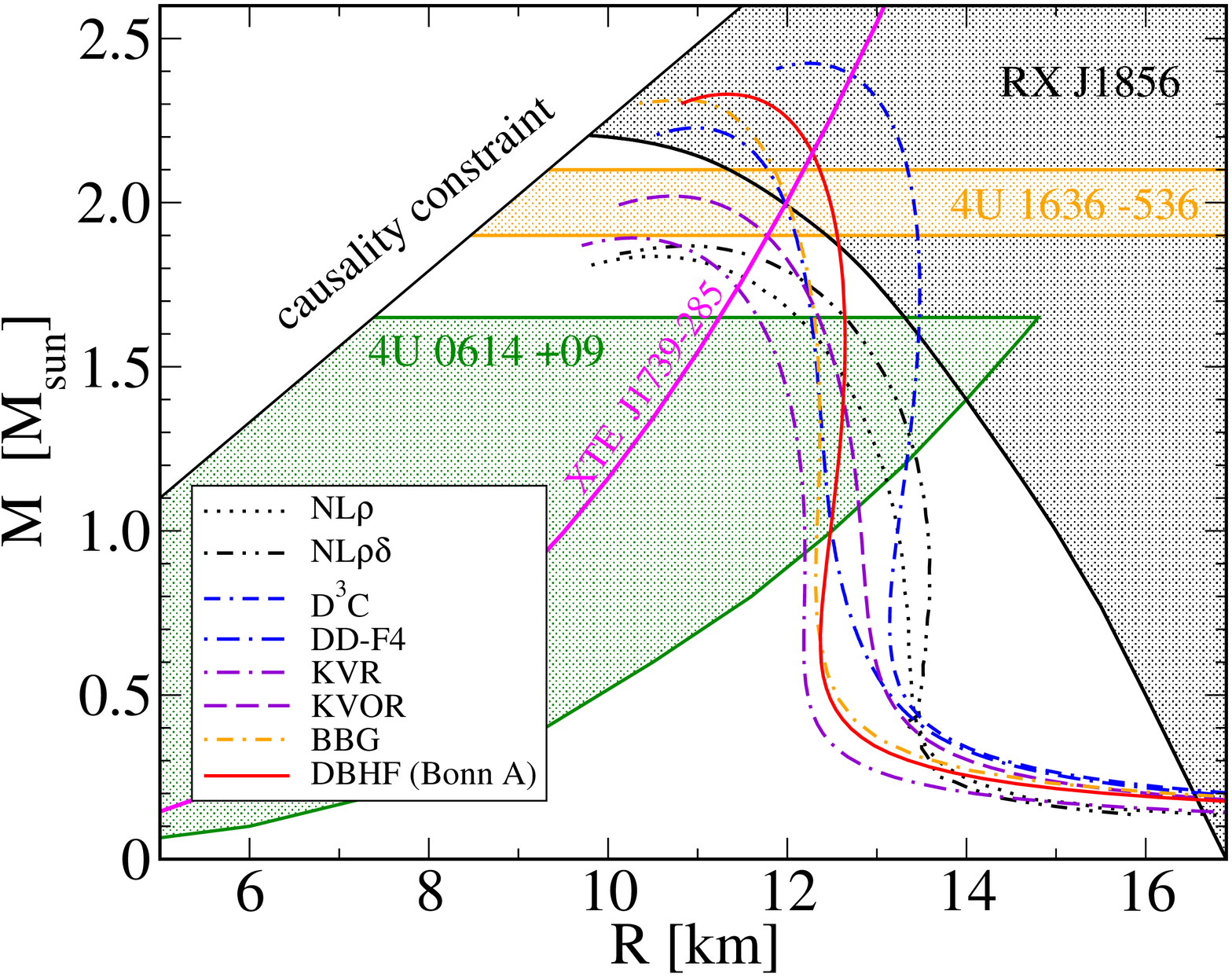}
\includegraphics[angle=270,width=0.5\textwidth]{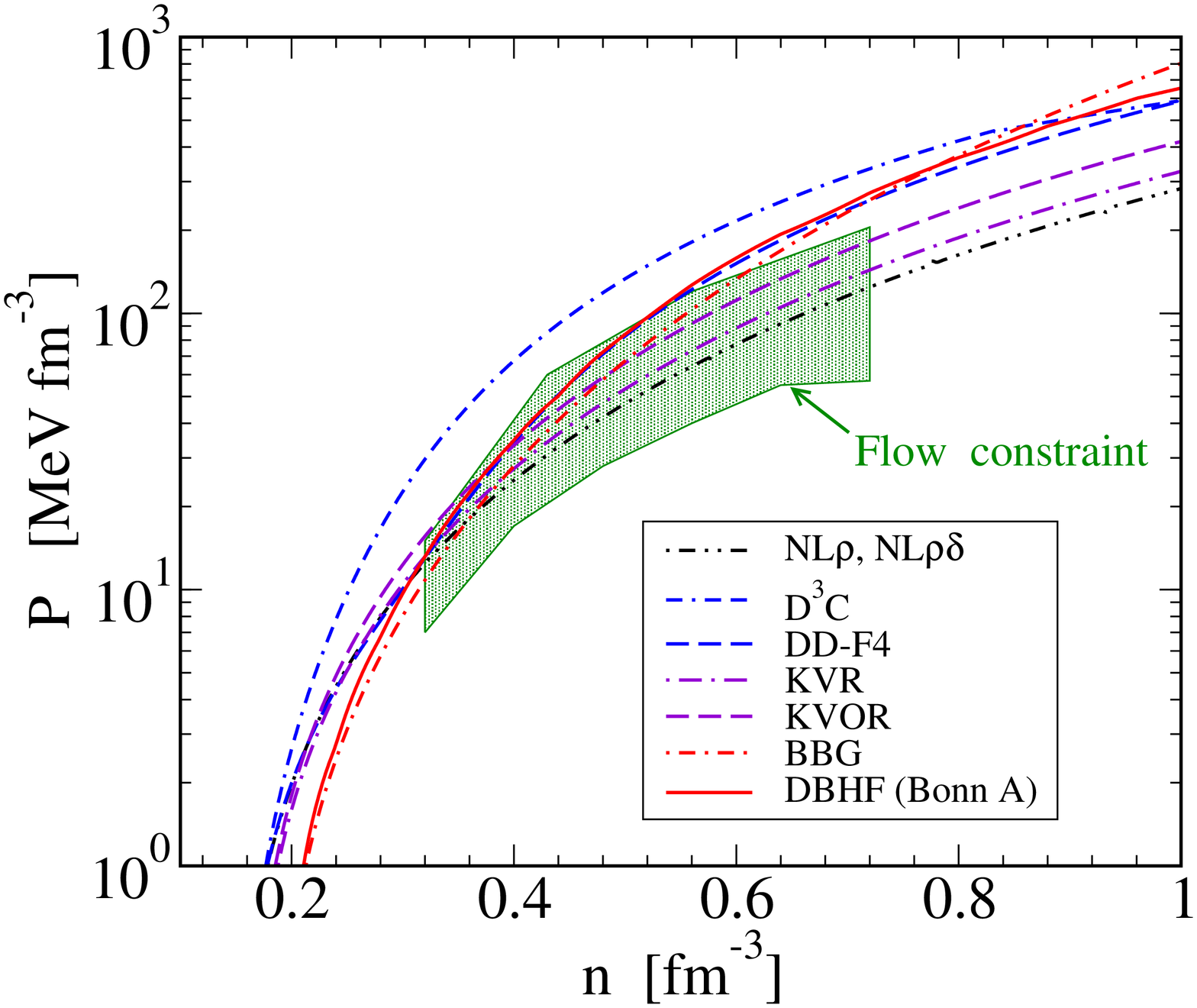}&
\end{tabular}
\caption{Left panel: Mass and mass-radius constraints on compact star 
configurations from recent observations compared to solutions of TOV equations 
for nuclear EoS discussed in the text.  
Right panel: Flow constraint from heavy-ion collisions 
\cite{Danielewicz:2002pu} compared to the same set of nuclear EoS used in the 
left panel. }
    \label{f:M-R}
\end{figure}

In the left panel of Fig.~\ref{f:M-R} we show some of these modern 
observational constraints for masses and mass-radius relationships together 
with solutions of the Tolman-Oppenheimer-Volkoff (TOV) equations for a set 
of eight hadronic EoS classified in three groups:  
(i) {\it relativistic mean-field (RMF) approaches} 
%\cite{Wal74,Ser86,Rei89,Rin96}
{\it with non-linear (NL) self-interactions} of the $\sigma$ 
meson \cite{Gaitanos:2003zg}. 
In NL$\rho$ the isovector part of the interaction is
described only by a $\rho$ meson, while the set NL$\rho\delta$ also includes a
scalar isovector meson $\delta$ that is usually neglected in RMF models
\cite{Liu02};
(ii) {\it RMF models with density dependent couplings and 
masses} 
are represented here by four different models from two classes, where in the 
first one density dependent meson couplings are modeled so that a number of  
properties of finite nuclei (binding energies, charge and diffraction radii, 
surface thicknesses, neutron skin in ${}^{208}$Pb, spin-orbit splittings) can 
be fitted \cite{Typel:2005ba}.
D${}^{3}$C has in addition a derivative coupling leading to momentum-dependent
nucleon self-energies and DD-F4 is modeled such that the flow constraint 
\cite{Danielewicz:2002pu} from heavy-ion collisions is fulfilled. 
The second class of these models is motivated by the Brown-Rho scaling 
assumption \cite{Brown:1991kk} that not only the nucleon mass but also the
meson masses should decrease with increasing density. 
In the KVR and KVOR models \cite{Kolomeitsev:2004ff}
these dependences were related to a 
nonlinear scaling function of the  $\sigma$- meson field such that the EoS of 
symmetric nuclear matter and pure neutron matter below four times the
saturation density coincide with those of the Urbana-Argonne group 
\cite{APR}.
In this way the latter approach builds a bridge between the phenomenological 
RMF models and (iii) {\it microscopic EoS} built on realistic 
nucleon-nucleon forces.
Besides the variational approaches (APR \cite{APR}, WFF \cite{Wiringa:1988tp},
FPS \cite{Friedman:1981qw}) such ab-initio approaches to 
nuclear matter are provided, e.g., by the relativistic
Dirac-\-Brueckner-\-Hartree-\-Fock (DBHF) \cite{DaFuFae04} 
and the nonrelativistic Brueckner-\-Bethe-\-Goldstone 
\cite{Baldo:1999rq} approaches.
Stiff EoS like D$^3$C, DD-F4, BBG and DBHF fulfill the demanding constraints
for a large radius and mass, while the softer ones like NL$\rho$ don't. 
It is interesting to note that the flow constraint \cite{Danielewicz:2002pu}
shown in the right panel of  Fig.~\ref{f:M-R} sets limits to the tolerable
stiffness: it excludes the D$^3$C EoS and demonstrates that DD-F4, BBG and DBHF
become too stiff at high densities above $\sim 0.55$ fm$^{-3}$.
For a detailed discussion, see Ref. \cite{Klahn:2006ir}.

A key question asked in investigating the structure of matter at high 
densities is how the quark substructure of hadrons manifests itself in the 
EoS and whether the phase transition to quark matter can occur inside compact 
stars. 
In Ref. \cite{Ozel:2006km}, \"Ozel has debated that the new %compact star 
constraints reported above would exclude quark matter in compact star 
interiors reasoning that it would entail an intolerable 
softening of the EoS. Alford et al. \cite{Alford:2006vz} have given a few 
counter examples demonstrating that quark matter cannot be excluded.
In the following section we discuss a recently developed chiral quark model 
\cite{Klahn:2006iw} which is in accord with the modern constraints, 
see also \cite{Blaschke:2007ri}.

\section{Color superconducting quark matter: masquerade revisited}

We describe the thermodynamics of the deconfined quark matter phase  
within a three-flavor quark model of Nambu--Jona-Lasinio (NJL) type, with a 
mean-field thermodynamic potential given by
\begin{eqnarray}  
\Omega_{MF}(T,\mu) 
%&=&       -\frac{1}{\beta V}\ln Z_{MF}(T,\mu)\nonumber\\  
&=& \frac{1}{8 G_S}\left[\sum_{i=u,d,s}(m^*_i-m_i)^2   
  - \frac{2}{\eta_V}(2\omega_0^2+\phi_0^2) 
        +\frac{2}{\eta_D}\sum_{A=2,5,7}|\Delta_{AA}|^2\right]   
        \nonumber \\  
&&-\int\frac{d^3p}{(2\pi)^3}\sum_{a=1}^{18}  
        \left[E_a+2T\ln\left(1+e^{-E_a/T}\right)\right]  
        + \Omega_l - \Omega_0~.  
\label{eos-blaschke-potential}  
\end{eqnarray}  
Here, $\Omega_l$ is the thermodynamic potential for electrons and muons,  
and the divergent term $\Omega_0$ is subtracted in order to assure zero  
pressure and energy density in vacuum ($T=\mu=0$). The quasiparticle  
dispersion relations, $E_a(p)$, are the eigenvalues of the hermitean matrix  
\begin{equation}  
{\mathcal M} = \left[  
\setlength\arraycolsep{-0.01cm}  
\begin{array}{cc}  
        -\vec{\gamma}\cdot\vec{p}- \hat{m}^*+\gamma^0\hat{\mu}^* &   
        i \gamma_5 C \tau_A \lambda_A \Delta_{AA} \\  
        i C \gamma_5 \tau_A \lambda_A\Delta_{AA}^* &   
        -\vec{\gamma}^T\cdot\vec{p}+\hat{m}^*-\gamma^0\hat{\mu}^*  
\end{array}  
\right],  
\label{eos-blaschke-eigmatrix}  
\end{equation}  
in color, flavor, Dirac, and Nambu-Gorkov space. Here, $\Delta_{AA}$ are the
diquark gaps. $\hat{m}^*$ is the diagonal renormalized mass matrix and
$\hat{\mu}^*$ the renormalized chemical potential matrix, $\hat{\mu}^*={\rm
  diag}_f
(\mu_u-G_S\eta_V\omega_0,\mu_d-G_S\eta_V\omega_0,\mu_s-G_S\eta_V\phi_0)$.  
The gaps and the renormalized masses are determined by minimization of the
mean-field thermodynamic potential (\ref{eos-blaschke-potential}).  
We have to obey constraints of charge neutrality which depend on the 
application we consider.
In the (approximately) isospin symmetric situation of a heavy-ion collision,
the color charges are neutralized, while the electric charge in general is
non-zero.  For matter in $\beta$-equilibrium in compact stars, also the 
global electric charge neutrality has to be fulfilled. For further details, see
\cite{Blaschke:2005uj,Ruster:2005jc,Abuki:2005ms,Warringa:2005jh}.

We consider $\eta_D$ as a free parameter of the quark matter model, to be
tuned with the present phenomenological constraints on the high-density EoS.
Similarly, the relation between the coupling in the scalar and vector meson
channels, $\eta_V$, is considered as a free parameter of the model.  The
remaining degrees of freedom are fixed according to the NJL model
parameterization in table I of \cite{Grigorian:2006qe}, where a fit to
low-energy phenomenological results has been made.

\begin{figure} [ht]
\begin{tabular}{ll}
\includegraphics[angle=270,width=0.5\textwidth]{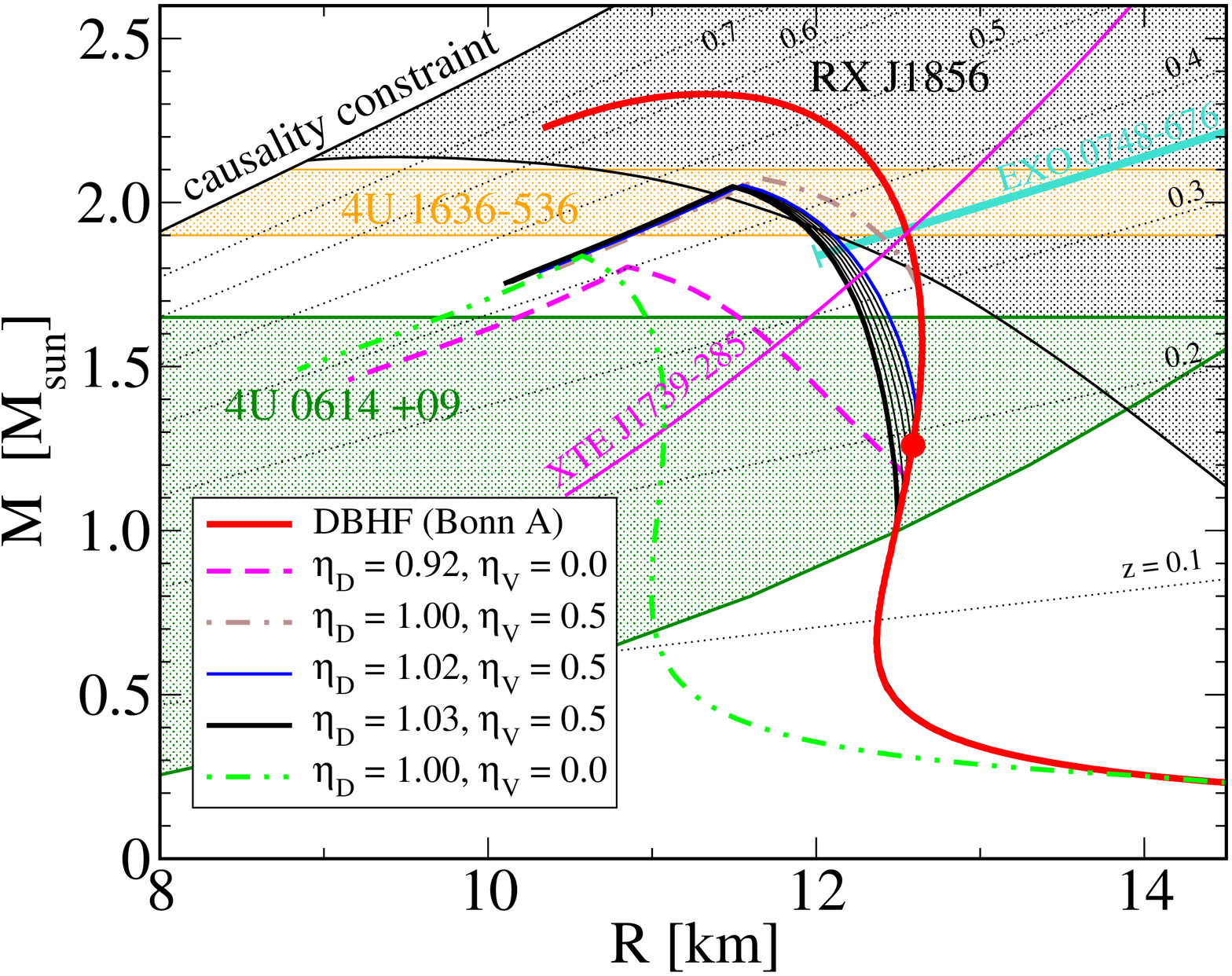}
\includegraphics[angle=270,width=0.5\textwidth]{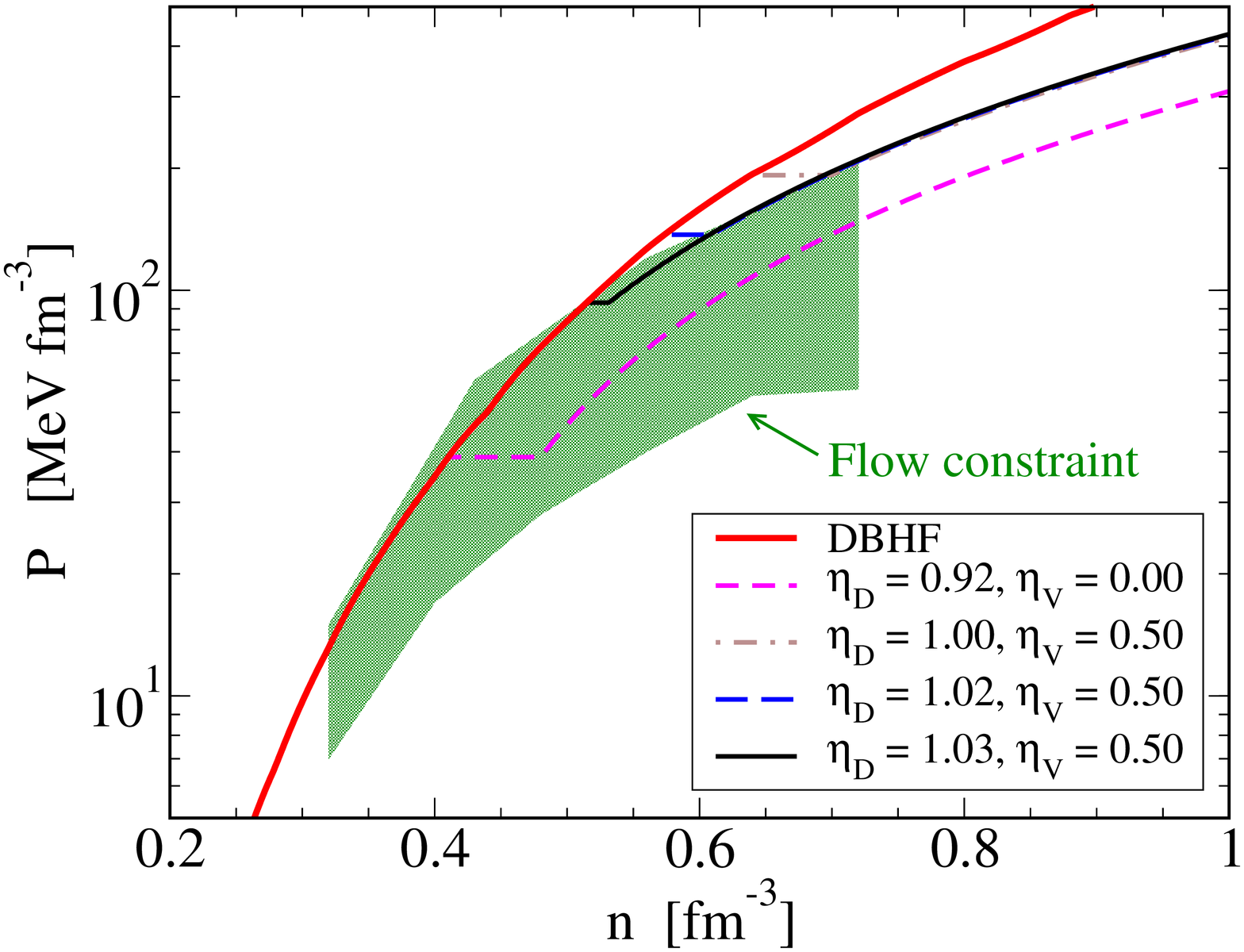}
\end{tabular}
\caption{Same as Fig.~\ref{f:M-R} for hybrid EoS  with a low density hadronic 
branch described by the DBHF approach and a high density quark matter branch 
obtained from a three-flavor NJL model with color superconductivity (diquark 
coupling $\eta_D$) and isoscalar vector meson coupling ($\eta_V$).}
    \label{f:P-n}
\end{figure}

As a unified description of quark-hadron matter, naturally
including a description of the phase transition, is not available yet, 
although steps in this direction have been suggested within nonrelativistic
\cite{Ropke:1986qs} and relativistic \cite{Lawley:2006ps} models. 
We apply here the so-called two-phase description, being aware of its
limitations. The nuclear matter phase is described within the
DBHF approach and the transition to the quark matter phase given above
is obtained by a Maxwell construction. 
In the right panel of Fig.~\ref{f:P-n}
it can be seen that the necessary softening of the high density EoS in 
accordance with the flow constraint is obtained for a vector coupling of 
$\eta_V=0.5$ whereas an appropriate deconfinement density is obtained for a 
strong diquark coupling in the range  $\eta_D=1.02 - 1.03$. The resulting phase
transition is weakly first order with an almost negligible density jump.
Applying this hybrid EoS with so defined free parameters under compact star
conditions a sequence of hybrid star configurations is obtained which fulfills
all modern constraints, see the left panel of Fig. 2. 
In that figure we also indicate by a red dot the minimal mass $M_{DU}$ for 
which the central density reaches a value allowing the fast direct Urca (DU) 
cooling process in DBHF neutron star matter to occur, leading to problems 
with cooling phenomenology \cite{Blaschke:2004vq,Blaschke:2006gd}.
Note that for a strong diquark coupling $\eta_D=1.03$, the critical density 
for quark deconfinement is low enough to prevent the hadronic direct Urca (DU) 
cooling problem by an early onset of quark matter. For the given hybrid 
EoS, there is a long sequence of stable hybrid stars with two-flavor 
superconducting (2SC) quark matter, before the occurrence of the strange quark 
flavor and the simultaneous onset of the color-flavor-locking (CFL) phase 
which renders the star gravitationally unstable 
\cite{Baldo:2002ju,Klahn:2006iw}.
Comparing the hybrid star sequences with the purely hadronic DBHF ones one can 
conclude that the former 'masquerade' themselves as neutron stars 
\cite{Alford:2004pf} by having very similar mechanical properties.

Besides for mass-radius relationships, the masquerade effect can be discussed 
also for the moment of inertia \cite{Klahn:2006iw}, which is becoming 
accessible to measurements in relativistic binary systems like the double 
pulsar J0737--3039, see \cite{Lattimer:2004nj,Bejger:2005jy}.
\begin{figure} [ht]
\includegraphics[width=7.5cm,height=14cm,angle=-90]{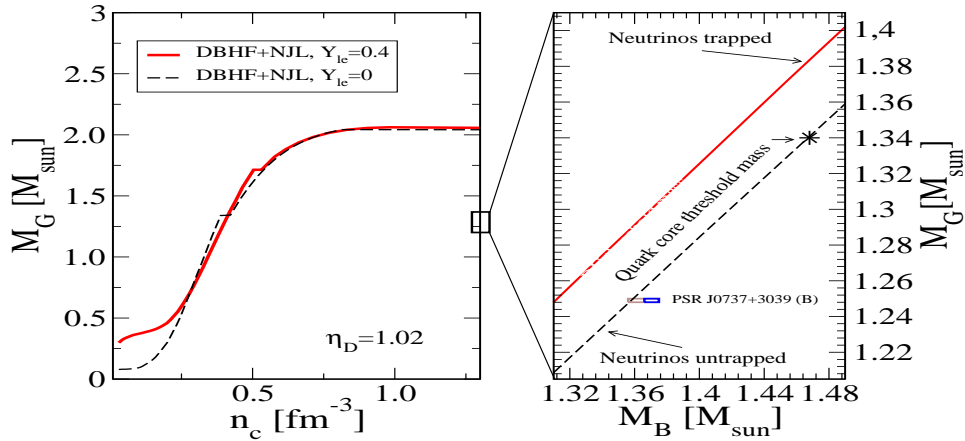}
\caption{The effect of neutrino untrapping on hybrid star configurations:
a mass defect of $0.04~M_\odot$ occurs in the transition from neutrino 
trapping case (solid red line, $Y_{le}=0.4$) to untrapping  
(dashed black lines, $Y_{le}=0$) at fixed baryon mass $M_B$, independent 
of the presence of a quark core for masses above the threshold indicated 
by an asterix. 
The blue rectangle corresponds to the constraint from the lighter 
companion (star B) in the double pulsar PSR J0737-3039, see \cite{Pod05}.  
}
\label{f:untrap}
\end{figure}

In these proceedings we add a new aspect to this masquerade discussion:the 
energy release in the neutrino untrapping transition of a cooling protoneutron 
star (PNS).
The process is pictured such that the PNS formed in a supernova collapse is 
hot enough ($T_{\rm PNS}\sim 30 \dots 50$ MeV) to trap neutrinos since their 
mean free path is shorter than the size of the star.
In this era the fast neutrino cooling cannot proceed from the volume but only 
from the surface. 
Until the neutrino opacity temperature ($T_{\rm opac} \sim 1$ MeV)
is reached, the neutrinos take part in the $\beta$-equilibration processes and 
a finite lepton fraction is established. We will assume $Y_{le}=0.4$ 
corresponding to a neutrino chemical potential $\mu_\nu\sim 200$ MeV in the 
PNS core. 
For a discussion of hot hybrid PNS, see \cite{Nicotra:2006eg}.
In the untrapping transition at $T\sim T_{\rm opac}$, the neutrinos 
decouple from the $\beta$-equilibrium so that $Y_{le}\to 0$ and the 
equilibrium configuration gets readjusted with a new gravitational mass while 
naturally, the baryon number is conserved in this process. 
From Fig. \ref{f:untrap} one can estimate the mass defect (energy release) due 
to neutrino untrapping to $\Delta M = 0.04~M_\odot$, almost independent of the 
mass of the configuration and even of its structure: neutron stars and hybrid 
stars result in the same energy release! 
This is another aspect of the masquerade effect. 
Note that the neutrino untrapped configurations fulfill the extended $M_G-M_B$
constraint \cite{Klahn:2006ir}, given by the rectangle in the right 
panel of Fig.\ref{f:untrap}. This constraint is derived from the lighter 
companion (star B) in the double pulsar J0737--3039 (B) \cite{Pod05}.

\section{Unmasking compact star interiors}
To unmask the neutron star interior 
and to disentangle a quark matter core from a hadronic one,
might therefore require observables 
based on transport properties, strongly modified from normal due to the color
superconductivity.
It has been suggested to base tests of the structure of matter at high 
densities on analyses of the cooling behavior 
\cite{Blaschke:2006gd,Popov:2004ey,Popov:2005xa}
or the stability of fastly rotating stars against r-modes 
\cite{Madsen:1999ci,Drago:2007iy}.
It has turned out that for these phenomena the fine tuning of color 
superconductivity in quark matter is an essential ingredient.
The result of these studies suggests that on the one hand unpaired quark matter
results in too fast cooling (same holds for pure two-flavor color 
superconductivity (2SC) due to one unpaired (blue) color) and on the other,
complete pairing with large gaps (CFL phase, $\Delta_{\rm CFL}\sim 100$ MeV)
would result in too slow cooling \cite{Page:2000wt,Blaschke:2000dy}, 
r-mode instability \cite{Madsen:1999ci} 
and gravitational collapse \cite{Baldo:2002ju,Klahn:2006iw}. 
Viable alternatives for quark matter thus have to be color superconducting,
with all quark modes paired, but a few modes (at least one) should have a 
small gap of $\Delta_X\sim 1$ MeV, desirable with a decreasing density 
dependence, as suggested for the 2SC+X phase 
\cite{Grigorian:2004jq,Grigorian:2006pu}.
Unfortunately, there is no microscopic model supporting the 2SC+X pairing 
pattern yet. A suitable candidate, however, could be the recently suggested 
single flavor pairing in the isotropic spin-1 color superconductivity phase
(iso-CSL) \cite{Aguilera:2005tg,Aguilera:2006cj}, see also
\cite{Marhauser:2006hy}. 
It has been shown that the neutrino emissivity and bulk viscosity 
of the iso-CSL phase fulfills constraints from cooling and r-mode stability 
\cite{Blaschke:2007bv}.

\section{Conclusions}
In this contribution it is shown that modern quantum field
theoretical approaches to quark matter including color
superconductivity and a vector meanfield allow a microscopic description of
hybrid stars which fulfill the new, strong constraints
for high masses, large radii, sufficiently fast cooling evolution and 
absence of r-mode instabilities.
The deconfinement transition in the resulting stiff hybrid equation of state 
is weakly first order so that signals of it have to be expected due to 
specific changes in transport properties governing the rotational and cooling 
evolution caused by the color superconductivity of quark matter. 
This conclusion holds analogously for the investigation of quark deconfinement 
in future generations of nucleus-nucleus collision experiments
at low temperatures and high baryon densities,
such as CBM @ FAIR.
The extrapolation of the hybrid EoS constrained here from compact star physics
results in almost crossover behavior at the deconfinement transition 
 \cite{Grigorian:2006pu}, so that promising observable effects for its 
diagnostics shall be based on (precursor) effects of color superconductivity.

\section*{Acknowledgements}
We thank the organizers of the EXOCT07 conference for providing a perfect 
environment for discussios of  exotic matter and compact star research.
D.B. is supported by the Polish Ministry of Science and Higher 
Education, T.K. is grateful for partial support from GSI Darmstadt and the
Department of Energy, Office of Nuclear Physics, contract no.\ 
DE-AC02-06CH11357.
F.S. acknowledges support from the Swedish Graduate   
School of Space Technology and the Royal Swedish Academy of Sciences.

\end{document}